\RequirePackage{ifpdf}
\ifpdf 
\documentclass[pdftex]{sigma}
\else
\documentclass{sigma}
\fi

\numberwithin{equation}{section}

\begin{document}

\allowdisplaybreaks
\renewcommand{\PaperNumber}{005}

\FirstPageHeading

\renewcommand{\thefootnote}{$\star$}

\ShortArticleName{Symmetry Operators for the FPK Equation with
Quadratic Nonlocal Nonlinearity}

\ArticleName{Symmetry Operators\\ for the  Fokker--Plank--Kolmogorov\\
Equation with Nonlocal Quadratic Nonlinearity\footnote{This paper
is a contribution to the Vadim Kuznetsov Memorial Issue
``Integrable Systems and Related Topics''. The full collection is
available at
\href{http://www.emis.de/journals/SIGMA/kuznetsov.html}{http://www.emis.de/journals/SIGMA/kuznetsov.html}}}

\Author{Alexander V. SHAPOVALOV~$^\dag$, Roman O. REZAEV~$^\ddag$
and Andrey Yu. TRIFONOV~$^\ddag$}

\AuthorNameForHeading{A.V. Shapovalov, R.O. Rezaev and A.Yu. Trifonov}

\Address{$^\dag$~Theoretical Physics Department, Tomsk State
University,\\
$\phantom{^\dag}$~36 Lenin Ave., 660050, Tomsk, Russia}
\EmailD{\href{mailto:shpv@phys.tsu.ru}{shpv@phys.tsu.ru}}
\Address{$^\ddag$~Laboratory of Mathematical Physics, Mathematical
Physics Department, \\
$\phantom{^\ddag}$~Tomsk Polytechnical  University, 30 Lenin Ave.,
660034, Tomsk, Russia}
\EmailD{\href{mailto:rezaev@tpu.ru}{rezaev@tpu.ru},
\href{mailto:trifonov@phtd.tpu.edu.ru}{trifonov@phtd.tpu.edu.ru}}

\ArticleDates{Received October 11, 2006, in f\/inal form December
09, 2006; Published online January 05, 2007}

\Abstract{The Cauchy problem for the Fokker--Plank--Kolmogorov
equation with a nonlocal nonlinear drift term is reduced to a
similar problem for the correspondent linear equation. The
relation between symmetry operators of the linear and nonlinear
Fokker--Plank--Kolmogorov equations is considered. Illustrative
examples of the one-dimensional symmetry operators are presented.}

\Keywords{symmetry operators; Fokker--Plank--Kolmogorov equation;
nonlinear partial dif\/ferential equations}

\Classification{35Q58; 37J15} 

\newtheorem{demo}{Statement}

\section{Introduction}

By def\/inition, symmetry operators leave invariant the solution
set of an equation and  allow to generate  new solutions from the
known ones  (see, e.g.~\cite{shapovalov:Miller,
shapovalov:MANKO}). The f\/inding of symmetry operators is an
important problem, but it rarely can be solved explicitly because
the equations that determine symmetry operators are complicated
and nonlinear. Therefore, special types of symmetry operators are
of interest. The most af\/fective approach is developed in the
framework of the group analysis of dif\/ferential equations
\cite{shapovalov:OVSIANNIKOV, shapovalov:IBRAGIMOV,
shapovalov:OLVER, shapovalov:FUSCH-SS, shapovalov:FUSCH-N} where
the Lie groups of symmetry operators are considered. The Lie group
generators (related to symmetries) are obtained from the
determining linear equations which can be solved in a regular way
when the symmetries are dif\/ferential operators. The symmetries
of dif\/ferential equations can be considered in the context of
dif\/ferential geometry \cite{shapovalov:OLVER,
shapovalov:KRASIL-LYCHAG-VINIGR, shapovalov:GAETA}.

The calculation of symmetries for integro-dif\/ferential equations
is a more complex problem because there are no general way  of
choosing an appropriate structure  of symmetries. The f\/inding of
symmetry operators for a nonlocal equation is usually a hopeless
task. Under these circumstances, examples of symmetry operators
for a nonlinear equation with nonlocal terms are of mathematical
interest.

In this work we consider an approach that can be used to obtain
such examples for the Fokker--Planck--Kolmogorov equation (FPKE)
of special form with a quadratic nonlocal non\-li\-nea\-rity
\begin{equation}
\Big \{-\partial_t+\varepsilon\Delta + \partial_{\vec x} \Big(\vec
V(\vec x,t)+\varkappa\int_{{\mathbb R}^n} \vec W(\vec
x,\vec y, t)u(\vec y,t)d\vec y \Big)\Big\} u(\vec x,t)=0,
\label{shapovalov:FPK-1}
\end{equation}
where
\begin{equation}
\vec V(\vec x,t)= K_1\vec x, \qquad  \vec W(\vec x, \vec y, t) =
K_2\vec x+K_3\vec y. \label{shapovalov:FPK-2}
\end{equation}
Here, $t\in \mathbb{R}^1$, $\vec x=(x_1,\ldots,x_n)^{\intercal}\in
\mathbb{R}^n$, $\vec y=(y_1,\ldots,y_n)^{\intercal}\in
\mathbb{R}^n$ are independent variables;
$(x_1,\ldots,x_n)^\intercal$ means a transpose to a vector or a
matrix; $d\vec x=dx_1\cdots dx_n$; the dependent variable $u(\vec
x,t)$ is a real smooth function decreasing as $\|\vec
x\|\to\infty$;  $K_1$, $K_2$, $K_3$ are arbitrary constant
matrices of order $n\times n$; $\varepsilon$ and $\varkappa$ are
real parameters; $\partial_t=\partial/\partial t$; $\partial_{\vec
x}=\partial/\partial \vec x$ is a gradient operator with respect
to $\vec{x}$; $\Delta=\partial_{\vec x}\partial_{\vec
x}=\sum\limits^n_{i=1}\partial^2/\partial x_i^2 $ is a Laplace
operator.

The operator of equation \eqref{shapovalov:FPK-1} is quadratic in
independent variables and in derivatives and it has a nonlocal
quadratic nonlinear term.

This equation serves as a simple example of  a class of
``near-linear'' nonlocal equations
\cite{shapovalov:BELUCH-Frank2005}, such that they are nonlinear
but the integrability problem for them can be reduced to seeking a
solution of  appropriate linear equations. Nonlinear equations of
such type regularly depend on the nonlinearity parameter and they
possess solutions which go into solutions of the linear equation
as the nonlinearity parameter tends to zero.

Equation \eqref{shapovalov:FPK-1} arises in mathematical problems
and it can be used in physical applications. In particular, the
FPKE \eqref{shapovalov:FPK-1},  \eqref{shapovalov:FPK-2} describes
the leading term of the asymptotic solutions constructed in
\cite{shapovalov:BELUCH-TRIFONOV} in the framework of the
formalism of semiclassical asymptotics \cite{shapovalov:MASLOV2,
shapovalov:BEL-DOB} for  equation~\eqref{shapovalov:FPK-1}, in
which $\vec V(\vec x,t)$ and $\vec W(\vec x,t)$ are real vector
functions of general form.

The semiclassical approximation is widely used in nonlinear
mathematical physics, providing a possibility of  constructing
explicit asymptotic solutions for  mathematical physics equations
coef\/f\/icients of which can be arbitrary smooth functions and
derivatives of dependent variables are assumed small. Most of the
equations solved by semiclassical methods are not exactly
integrable. For these equations, semiclassical methods of\/fer a
unique opportunity to investigate them analytically.

A method of semiclassical asymptotics based on the formalism of
the Maslov complex germ \cite{shapovalov:MASLOV2,
shapovalov:BEL-DOB, shapovalov:Dobrokh-Martin,
shapovalov:Albeverio, shapovalov:maslovv} has been developed for a
many-dimensional nonstationary Hartree type equation with nonlocal
nonlinearity in a class of functions localized in a neighborhood
of some phase curve \cite{shapovalov:BEL-SHAP-TRIF1,
shapovalov:BEL-SHAP-TRIF2, shapovalov:LIS-TRIF-SHAP1,
shapovalov:LIS-TRIF-SHAP3}. This class of functions has been
called the class of trajectory-concentrated functions (TCF). The
Hartree type equation whose operator is quadratic in independent
variables and derivatives provides another example of the class of
``near-linear'' nonlinear equations similar to the  FPKE
\eqref{shapovalov:FPK-1}.

The symmetry analysis area may be augmented by the study of the
symmetry features of the semiclassical approximation  as the
semiclassical methods are hoped to result in a new kind of
symmetries for mathematical physics equations.  The group
properties of the semiclassical approximation were considered in
quantum mechanics and in some models of the quantum f\/ield theory
\cite{shapovalov:SHVEDOV}. The semiclassical method for solving
the Cauchy problem  in the class of TCF's has been  developed for
the Hartree type equation \cite{shapovalov:BEL-SHAP-TRIF1,
shapovalov:BEL-SHAP-TRIF2, shapovalov:LIS-TRIF-SHAP1,
shapovalov:LIS-TRIF-SHAP3} and
 for the one-dimensional FPKE \cite{shapovalov:BELUCH-TRIFONOV, shapovalov:TRIF2,
shapovalov:TRIF-REZAEV_izv}. For the Hatree type equation,
symmetry operators have been found in the TCF class.

A nonlinear FPKE was used to analyze  stochastic processes in
various physical phenomena. In this connection,  the following
works where the Fokker--Plank--Kolmo\-go\-rov equation with the
nonlinear drift term  similar to that in \eqref{shapovalov:FPK-1}
was considered deserve mention (see also
\cite{shapovalov:BELUCH-Frank2005} and references therein). 
M.~Shiino and K.~Yoshida studied noise ef\/fects and phase 
transitions ef\/fects
involving chaos-nonchaos bifurcations
\cite{shapovalov:SHIINO-2001} in the framework of nonlinear
Fokker--Plank equations. These equations are shown to exhibit the
property of  $H$ theorem with a Lyapunov functional that takes the
form of free energy involving generalized entropies of Tsallis
\cite{shapovalov:SHIINO-2003}. In
\cite{shapovalov:DROZDOV-MORILLO-1} the stochastic resonance
phenomenon is discussed; in \cite{shapovalov:SCHULLER-VOGT} binary
branching and dying processes were studied.  The evolution of
quantum systems was described by means of  nonlinear 
FPKE's~\cite{shapovalov:FRANK-2004} where the nonlinearity ref\/lects the
quantum constraints imposed by the Bose and Fermi statistics.

The paper is organized as follows. In
Section~\ref{shapovalov:SEC-CAUCHY} the nonlinear FPKE is
presented with ne\-ces\-sary notations and def\/initions. The
Cauchy problem for the nonlinear FPKE is reduced to a~similar
problem for the corresponding linear FPKE in the class of
functions decreasing at inf\/inity via the Cauchy problem for the
f\/irst moment vector of a solution of the nonlinear FPKE. With
the help of the Cauchy problem solution we construct a nonlinear
evolution operator and the corresponding  left inverse operator in
explicit form  for the nonlinear FPKE.

In Section~\ref{shapovalov:SEC-SYMM} a general class of nonlinear
symmetry operators is considered for the non\-li\-near FPKE. The
symmetry operators are introduced in dif\/ferent ways, in
particular using the evolution operator and the left inverse
operator. Examples of one-dimensional symmetry ope\-ra\-tors are
given in explicit form as an illustration. In Conclusion the
results are discussed in the framework of  symmetry analysis.

\section{The Cauchy problem and the evolution operator}
\label{shapovalov:SEC-CAUCHY}

In our consideration, the  key part is   played by the Cauchy
problem for the FPKE \eqref{shapovalov:FPK-1},
\eqref{shapovalov:FPK-2} in the class of functions $u(\vec x,t)$
decreasing as $\|\vec x\|\to\infty$ at every point of time
$t\geqslant 0$. To be specif\/ic,  we assume that $u(\vec x,t)$
belongs to the Schwartz space $\mathcal{S}$ in the variable $\vec
x\in{\mathbb R}^n$ and regularly depends on $t$, i.e.  $u(\vec
x,t)$ is expanded as a power series in $t$ about $t=0$. Obviously,
equation~(\ref{shapovalov:FPK-1}) can be written in the form of
the balance equation
\begin{equation*}
\partial_t u(\vec x,t)=\partial_{\vec x}\vec B(\vec x,t,u),
\end{equation*}
where \[\vec B(\vec x,t,u)=\varepsilon \partial_{\vec x}u(\vec
x,t)+\vec V(\vec x,t)u(\vec x,t)+\varkappa\int_{{\mathbb R}^n}
\vec W(\vec x,\vec y, t)u(\vec y,t)d\vec y u(\vec x,t).\] Then
according to the divergence theorem, we obtain that the integral
$\int_{\mathbb{R}^n}u(\vec x,t)d\vec x$ conserves in time $t$ for
every solution $u(\vec x, t)$ of equation
\eqref{shapovalov:FPK-1}. Therefore, taking the initial function
$u(\vec x,0)=\gamma(\vec x)$ to be normalized,
$\int_{\mathbb{R}^n}\gamma(\vec x)d\vec x=1$, we can assume
\begin{equation}
\label{shapovalov:NORM} \int_{\mathbb{R}^n}u(\vec x,t)d\vec x=1,
\qquad t\geqslant 0,
\end{equation}
without loss of generality. We do not pay special attention to the
positive def\/initeness of solutions of the FPKE
\eqref{shapovalov:FPK-1},  leaving this requirement for specif\/ic
examples (see \cite{shapovalov:TATARSKY} for details).

Let us write equations  \eqref{shapovalov:FPK-1},
\eqref{shapovalov:FPK-2} in equivalent form
\begin{equation}\label{shapovalov:FPKE-1}
\{-\partial_t+\widehat H_{\rm nl}(\vec x,t;\vec X_u(t))\}u(\vec
x,t)=0,
\end{equation}
where the operator $\widehat H_{\rm nl}$ reads
\begin{equation}\label{shapovalov:FPKE-1a}
\widehat H_{\rm nl}(\vec x,t;\vec X_u(t))=\varepsilon\Delta +
\partial_{\vec x}\big(\Lambda\vec x+\varkappa K_3\vec X_u(t)\big),
\end{equation}
the matrix $\Lambda$ is
\begin{equation*}
\Lambda=K_1+\varkappa K_2,
\end{equation*}
and the vector
\begin{equation}\label{shapovalov:VECTOR-X}
\vec X_u(t)=\int_{{\mathbb R}^n}\vec xu(\vec x,t)d\vec x
\end{equation}
is the f\/irst moment of the function $u(\vec x,t)$. With the
obvious notation $\dot{\vec X}_u(t)=d\vec X_u(t)/d t$, we obtain
immediately from
\eqref{shapovalov:FPKE-1}--\eqref{shapovalov:VECTOR-X}, and
\eqref{shapovalov:NORM}
\begin{equation}\label{shapovalov:EE}
\dot{\vec X}_u(t)=-(\Lambda+\varkappa K_3)\vec X_u(t).
\end{equation}
Equation  \eqref{shapovalov:EE} can be considered  the f\/irst
equation of the Einstein--Ehrenfest system (EES) that describes
the evolution of the moments and centered high-order moments of a
solution $u(\vec x,t)$ of the FPKE \eqref{shapovalov:FPK-1} with
the vector functions $\vec V(\vec x,t)$ and $\vec W(\vec x,t)$ of
general form. The total EES for moments of all orders was derived
in constructing  approximate semiclassical solutions for a
one-dimensional  FPKE in \cite{shapovalov:BELUCH-TRIFONOV}.

\subsection{Solution of the Cauchy problem}

Let us set the Cauchy problem for equation
\eqref{shapovalov:FPKE-1}:
\begin{equation}\label{shapovalov:CAUCHY}
u(\vec x,0)=\gamma(\vec x),\qquad \gamma(\vec x)\in \mathcal{S},
\qquad\int_{\mathbb{R}^n} \gamma(\vec x)d\vec x=1.
\end{equation}

Then  we have the induced Cauchy problem for the vector $\vec
X_u(t)$
\begin{equation}\label{shapovalov:CAUCHY-X}
\vec X_{u}(0)=\vec X_\gamma= \int_{\mathbb{R}^n}\vec x\gamma(\vec
x)d\vec x
\end{equation}
determined by  \eqref{shapovalov:EE}.

The nonlinear Cauchy problem~\eqref{shapovalov:FPKE-1},
\eqref{shapovalov:CAUCHY} is reduced to a linear one as follows.
For a~given initial function $\gamma(\vec x)$
\eqref{shapovalov:CAUCHY}, we can seek a solution of the Cauchy
problem \eqref{shapovalov:EE}, \eqref{shapovalov:CAUCHY-X}
independently of the solution of equation
\eqref{shapovalov:FPKE-1} and obtain the  vector $\vec X_u(t)$
having the form of~\eqref{shapovalov:VECTOR-X} due to the
uniqueness of the Cauchy problem solution. Let us introduce a
function $w(\vec x,t)$ by the equality
\begin{equation}\label{shapovalov:W}
u(\vec x,t)=w(\vec x-\vec X_u(t),t).
\end{equation}
By substitution of \eqref{shapovalov:W} in
\eqref{shapovalov:FPKE-1} we obtain for the function $w(\vec x,t)$
a linear equation:
\begin{gather}
 -\partial_t w(\vec x,t)+\widehat Lw(\vec x,t)=0,\label{shapovalov:LIN-EQ-1} \\
\widehat L=\varepsilon\Delta+\partial_{\vec x}\Lambda\vec x.
 \label{shapovalov:LIN-EQ-2}
\end{gather}

From \eqref{shapovalov:CAUCHY} and \eqref{shapovalov:W} we have
\begin{equation}\label{shapovalov:LIN-EQ-3}
w(\vec x,0)=\tilde\gamma(\vec x)=\gamma(\vec x+\vec X_\gamma)
\end{equation}
and
\begin{equation}\label{shapovalov:LIN-EQ-3a}
\int_{\mathbb{ R}^n} \tilde \gamma(\vec x)d\vec x=1.
\end{equation}
Equation  \eqref{shapovalov:CAUCHY-X} results in
\begin{equation}\label{shapovalov:LIN-EQ-4}
\int_{\mathbb{ R}^n} \vec x\tilde\gamma(\vec x)d\vec x=0,
\end{equation}
i.e.\ the function $\tilde\gamma(\vec x)$ is centered. Obviously,
the integral $\int_{{\mathbb R}^n}w(\vec x,t)d\vec x$ conserves in
time $t$ for any solution $w(\vec x,t)$ of equation
\eqref{shapovalov:LIN-EQ-1}; then from
\eqref{shapovalov:LIN-EQ-3a} we have
\begin{equation*}
\int_{\mathbb{ R}^n} w(\vec y, t)d\vec y=1.
\end{equation*}

Therefore, the nonlinear Cauchy problem \eqref{shapovalov:FPKE-1},
\eqref{shapovalov:FPKE-1a},  \eqref{shapovalov:CAUCHY} can be
solved  as follows. First, for a given initial function
$\gamma(\vec x)$ \eqref{shapovalov:CAUCHY}  we solve the linear
Cauchy problem \eqref{shapovalov:LIN-EQ-1},
\eqref{shapovalov:LIN-EQ-2}, \eqref{shapovalov:LIN-EQ-3} with the
initial function $\tilde\gamma(\vec x)$ normalized by condition
\eqref{shapovalov:LIN-EQ-3a} and centered by
\eqref{shapovalov:LIN-EQ-4}. Second, we f\/ind the vector $\vec
X_u(t)$ by solving the Cauchy problem \eqref{shapovalov:EE},
\eqref{shapovalov:CAUCHY-X}. Then, the solution of the nonlinear
Cauchy problem \eqref{shapovalov:FPKE-1},
\eqref{shapovalov:FPKE-1a},  \eqref{shapovalov:CAUCHY} is given by
\eqref{shapovalov:W}.

Equation \eqref{shapovalov:LIN-EQ-1} is known (see, e.g.,
\cite{shapovalov:MANKO}) to have a solution in the form of a
Gaussian wave packet:
\begin{equation}\label{shapovalov:GAUSS-1}
w(\vec x,t)=\sqrt{\frac{\det Q(t)}{(2\pi \varepsilon)^n}}
\exp\Big[-\frac{1}{2\varepsilon}\vec x^\intercal Q(t)\vec x \Big],
\end{equation}
where $Q(t)$ is a symmetric positive-def\/inite matrix of  order
$n\times n$. Substituting \eqref{shapovalov:GAUSS-1} in
\eqref{shapovalov:LIN-EQ-1}, we obtain
\begin{gather*}
\vec x^\intercal\dot Q(t)\vec x+2\,{\vec x^\intercal (Q(t))^2\vec
x}-\vec x^\intercal \Lambda^\intercal Q(t)\vec x-
\vec x^\intercal Q(t)\Lambda\vec x-\varepsilon\displaystyle\frac{d}{d t}\log\det{Q(t)}\nonumber\\
\qquad{}+2\varepsilon{\rm Tr}\,(-\,Q(t)+\Lambda)
 =0.
\end{gather*}
Here $\mathop{\rm Tr}\nolimits\Lambda$  is the trace of the matrix
$\Lambda$. Equating the coef\/f\/icients of equal powers of $\vec
x$, we have
\begin{gather}
 \dot Q(t)+2(Q(t))^2-\Lambda^\intercal Q(t)-Q(t)\Lambda=0,\label{shapovalov:GAUSS-2}\\
-\frac{d}{d t}\log\det{Q(t)}+2\,{\rm Tr}\,(-\,Q(t)+\Lambda)=0. \nonumber 
\end{gather}
Let us  take $Q(t)$ in the form
\begin{equation}
Q(t)=B(t)(C(t))^{-1},\label{shapovalov:GAUSS-4}
\end{equation}
where  $B(t)$ and $C(t)$ are matrices of  order $n\times n$. On
substitution of \eqref{shapovalov:GAUSS-4} in
\eqref{shapovalov:GAUSS-2} we can write
\begin{alignat}{3}
    & \dot B(t)=\Lambda^\intercal B(t),\qquad && B(0)=B_0,\nonumber \\
    & \dot C(t)=2B(t)-\Lambda C(t),\qquad && C(0)=C_0,&
\label{shapovalov:SYST-VAR}
\end{alignat}
where $B_0$ and $C_0$ are arbitrary constant matrices of $n\times
n$ order. We call equations  \eqref{shapovalov:SYST-VAR} a~system
in variations in matrix form.

For the one-dimensional case, the  linear equation
\eqref{shapovalov:LIN-EQ-1} takes the form
\begin{equation}
\{-\partial_t+\varepsilon\partial_x^2+\partial_x\Lambda
x\}w(x,t)=0, \label{shapovalov:LIN-EQ-1D}
\end{equation}
where $\partial_x=\partial/\partial x$.

The solution \eqref{shapovalov:GAUSS-1} of equation
\eqref{shapovalov:LIN-EQ-1D} reads
\begin{equation*}
w (x,t)=\sqrt{\frac{B(t)}{2\pi\varepsilon C(t)}}
\exp\left[-\frac{B(t)}{2\varepsilon C(t)}x^2\right],
\end{equation*}
where  $B(t)$ and $C(t)$ are a solution of the system in
variations \eqref{shapovalov:SYST-VAR} in the one-dimensional
case. For $t=0$ we have
\begin{equation}
w(x,0)=\widetilde{\gamma}(x)=\sqrt{\frac{B_0}{2\pi\varepsilon
C_0}} \exp\left[-\frac{B_0}{2\varepsilon C_0}x^2\right].
\label{shapovalov:GAUSS-1D-1}
\end{equation}
Notice that  the function $\widetilde{\gamma}(x)$
\eqref{shapovalov:GAUSS-1D-1} is normalized and centered:
\[
\int_{-\infty}^{+\infty}\tilde{\gamma}(x)dx=1, \qquad
\int_{-\infty}^{+\infty}x\tilde{\gamma}(x)dx=0.
\]

Then the function
\begin{equation}
u(x,t)=w(x-X_{u}(t),t)=\sqrt{\frac{B(t)}{2\pi\varepsilon C(t)}}
\exp\left[-\frac{B(t)}{2\varepsilon C(t)} (x-X_u(t))^2\right]
\label{shapovalov:GAUSS-1D-2}
\end{equation}
will be a solution of the nonlinear equation
\begin{equation}
\{-\partial_t+\varepsilon\partial_x^2+\partial_x\Lambda
x+\varkappa K_3
X_u(t)\partial_x\}u(x,t)=0,\label{shapovalov:GAUSS-1D-3}
\end{equation}
with the initial condition
\begin{equation}
u(x,0)=\gamma(x)=\tilde{\gamma}(x-X_{\gamma})=\sqrt{\frac{B_0}{2\pi\varepsilon
C_0}} \exp\left[-\frac{B_0}{2\varepsilon
C_0}(x-X_{\gamma})^2\right].\label{shapovalov:GAUSS-1D-4}
\end{equation}

The vector $X_u(t)=\int_{-\infty}^{+\infty}xu(x,t)dx$ in equation
\eqref{shapovalov:GAUSS-1D-3} satisf\/ies the condition
\begin{equation}
\dot X_u(t)=-(\Lambda+\varkappa K_3)X_u(t),\qquad
X_u(0)=X_{\gamma}.\label{shapovalov:GAUSS-1D-5}
\end{equation}

\subsection{The evolution operator for a nonlinear FPKE}

Let us rewrite  the solution of the above nonlinear Cauchy problem
in terms of the corresponding nonlinear evolution operator.

Let $G_{\rm lin}(t,s,\vec x,\vec y)$  be the Green function of the
linear equation \eqref{shapovalov:LIN-EQ-1}, i.e.
\begin{equation*}
w(\vec x,t)=\int_{\mathbb{R}^n}G_{\rm lin}(t,s, \vec x,\vec y)
\tilde\gamma(\vec y)d\vec y.
\end{equation*}
Substituting $\vec x$ for  $\vec x-\vec X_u(t)$, we f\/ind
\begin{equation*}
w(\vec x-\vec X_u(t),t)=\int_{\mathbb{R}^n}G_{\rm lin}(t,s, \vec
x-\vec X_u(t),\vec y) \gamma(\vec y +\vec X_\gamma)d\vec y .
\end{equation*}
According to \eqref{shapovalov:W} and \eqref{shapovalov:LIN-EQ-3},
the function
\begin{equation}\label{shapovalov:GREEN-NLIN}
u(\vec x,t)=\int_{\mathbb{R}^n}G_{\rm nl}(t,s, \vec x,\vec
y,\gamma) \gamma(\vec y)d\vec y= \int_{\mathbb{R}^n}G_{\rm lin}
(t,s,\vec x-\vec X_u(t),\vec y-\vec X_\gamma) \gamma(\vec y)d\vec
y
\end{equation}
is a solution of the nonlinear equation \eqref{shapovalov:FPKE-1},
\eqref{shapovalov:FPKE-1a} with the initial condition
\eqref{shapovalov:CAUCHY}. Therefore,
\begin{equation}\label{shapovalov:GREEN-NLIN1}
G_{\rm nl}(t,s,\vec x,\vec y,\gamma)=G_{\rm lin}(t,s,\vec x-\vec
X_u(t),\vec y-\vec X_\gamma)
\end{equation}
is the  kernel of the evolution operator for the nonlinear
equation \eqref{shapovalov:FPKE-1}. Here $\gamma(\vec x)=u(\vec
x,s)$ and the initial time $t=0$ is replaced by $t=s$.

Suppose that   a solution of the system in variations
\eqref{shapovalov:SYST-VAR} has the form
\begin{gather}
B(t)=M_1(t,s)B_0,\qquad
C(t)=M_2(t,s)B_0+M_3(t,s)C_0,\label{shapovalov:SYST-IN-VAR}
\end{gather}
where $M_1(t,s)$, $M_2(t,s)$, and $M_3(t,s)$ are the matrix blocks
of order $n\times n$ of the matriciant (evolution matrix) of the
system in variations \eqref{shapovalov:SYST-VAR}. These matrix
blocks must satisfy the  condition
\begin{equation}
\dot{\rm M}={\rm AM}, \qquad {\rm M}(s) =\mathbb{I}_{2n\times
2n},\label{shapovalov:MATRICIANT}
\end{equation}
where
\[ {\rm M}={\rm M}(t,s)=\begin{pmatrix} M_1(t,s) & 0\\ M_2(t,s)&
M_3(t,s)\end{pmatrix},\qquad {\rm A}=\begin{pmatrix}
\Lambda^\intercal & 0\\2\,\mathbb{I}_{n\times n}& -\Lambda
\end{pmatrix}. \]
Set the Cauchy problem for \eqref{shapovalov:SYST-IN-VAR}:
\[
B(s)=B_0^\intercal=B_0,\qquad C(s)=C_0=0.
\]
The  Green's function of equation  \eqref{shapovalov:LIN-EQ-1}  is
known and can be taken as (see, e.g., \cite{shapovalov:MANKO}):
\begin{gather*}
 G_{\rm lin}(t,s,\vec{x},\vec{y})=\frac{1}{\sqrt{(2\pi\varepsilon)^n
  \det\big[M_2(t,s)\big(M_1(t,s)\big)^{-1}\big]}}\\
\phantom{G_{\rm lin}(t,s,\vec{x},\vec{y})=}{}
\times\exp\left\{-\frac{1}{2\varepsilon}
  (\vec{x}-M_3(t,s)\vec{y})^\intercal[M_1(t,s)(M_2(t,s))^{-1}]
  (\vec{x}- M_3(t,s)\vec{y})\right\}.\nonumber
\end{gather*}
Then the nonlinear evolution $\widehat U(t,s,\cdot)$ operator
(\ref{shapovalov:GREEN-NLIN}) reads
\begin{equation}
\widehat U(t,s,\gamma)(\vec x)=\displaystyle\int_{{\mathbb
R}^n}G_{\rm nl}(t,s,\vec x,\vec y ,\gamma)\gamma(\vec y)d\vec y.
\label{shapovalov:NL-EVOL}
\end{equation}

The left-inverse operator $\widehat U^{-1}(t,s,\cdot)$ for the
operator \eqref{shapovalov:NL-EVOL} is
\begin{equation}
\widehat U^{-1}(t,s,u)(\vec x)=\int_{{\mathbb R}^n}G^{-1}_{\rm
nl}(t,s,\vec x,\vec y,u)u(\vec x,t)d\vec x.
\label{shapovalov:NL-EVOL-1}
\end{equation}
Here  $G^{-1}_{\rm nl}(t,s,\vec x,\vec y,u)$ is the  kernel of the
left-inverse operator, which is obtained from
\eqref{shapovalov:GREEN-NLIN1} if we substitute $t$ for  $s$ and
$s$ for $t$. The explicit form of this function is
\begin{gather*}
G^{-1}_{\rm nl}(t,s,\vec{x},\vec{y},u)=\frac{1}
{\sqrt{{(2\pi\varepsilon)^n\det\big[M_2(s,t)\big(M_1(s,t)\big)^{-1}\big]}}}\nonumber
\\
\phantom{G^{-1}_{\rm nl}(t,s,\vec{x},\vec{y},u)=}{}
\times\exp\Big\{-\frac{1}{2\varepsilon}\big(\vec{x}-\vec
X_{\gamma}-
  M_3(s,t)(\vec{y}-\vec X_u(t))\big)^\intercal\big[M_1(s,t)(M_2(s,t))^{-1}\big]\nonumber\\
\phantom{G^{-1}_{\rm nl}(t,s,\vec{x},\vec{y},u)=}{}
\times\big(\vec{x}-\vec X_{\gamma}-M_3(s,t) (\vec{y}-\vec
X_u(t))\big)\Big\}.
\end{gather*}

The Green's function for  the one-dimensional equation
\eqref{shapovalov:LIN-EQ-1D} is
\begin{equation*}
G_{\rm lin}(t,s,x,y)=\sqrt{\frac{M_1(t,s)}{2\pi \varepsilon
M_2(t,s)}} \exp\left[-\frac{1}{2\varepsilon}
\frac{M_1(t,s)}{M_2(t,s)}(x-M_3(t,s)y)^2\right],
\end{equation*}
where $M_1(t,s)$, $M_2(t,s)$, $M_3(t,s)$ are solution of the
system in variation
 \eqref{shapovalov:MATRICIANT} in the one-dimensional case.

The kernel of the evolution operator for the nonlinear equation
\eqref{shapovalov:GAUSS-1D-3} takes the form
\begin{gather}
G_{\rm nl}(t,s, x,y,\gamma)=G_{\rm lin}(t,s,x-X_u(t),y-X_{\gamma})\label{shapovalov:REZ8}\\
\phantom{G_{\rm nl}(t,s,
x,y,\gamma)}{}=\sqrt{\frac{M_1(t,s)}{2\pi\varepsilon M_2(t,s)}}
  \exp\left[-\frac{1}{2\varepsilon}\frac{M_1(t,s)}{M_2(t,s)}
  \big(x-X_u(t)-M_3(t,s)(y-X_{\gamma})\big)^2\right],\nonumber
\end{gather}
where $X_u(t)$ satisf\/ies equation~\eqref{shapovalov:GAUSS-1D-5}.
The evolution operator~\eqref{shapovalov:NL-EVOL} with the
kernel~\eqref{shapovalov:REZ8} is written as{\samepage
\begin{equation}
u(x,t)=\widehat U(t,s,\gamma)(x)=\int_{-\infty}^{+\infty} G_{\rm
lin}(t,s,x-X_u(t),y-X_{\gamma})\gamma(y)dy.
\label{shapovalov:REZ9}
\end{equation}
Here the function $u(x,t)$ having  the form of
\eqref{shapovalov:REZ9} is a solution of equation
\eqref{shapovalov:GAUSS-1D-3}.}

Notice that direct calculation of the action of the evolution
operator \eqref{shapovalov:NL-EVOL} on the function~$\gamma(y)$
having the form of \eqref{shapovalov:GAUSS-1D-4} gives  the
function \eqref{shapovalov:GAUSS-1D-2}:
\begin{gather*}
\widehat U(t,s,\gamma)(\vec
x)=\int_{-\infty}^{+\infty}\sqrt{\frac{M_1(t,s)}{2\pi\varepsilon
  M_2(t,s)}}\\
  \phantom{\widehat U(t,s,\gamma)(\vec x)=}{}\times \exp\left[-\frac{1}{2\varepsilon}\frac{M_1(t,s)}{M_2(t,s)}
  \big(x-X_u(t)-M_3(t,s)(y-X_{\gamma})\big)^2\right]\\
\phantom{\widehat U(t,s,\gamma)(\vec x)=}{} \times
  \sqrt{\frac{B_0}{2\pi\varepsilon C_0}}\exp\left[-\frac{B_0}{2\varepsilon
  C_0}(y-X_{\gamma})^2\right]dy\\
\phantom{\widehat U(t,s,\gamma)(\vec x)=}{}
  =\sqrt{\frac{B(t)}{2\pi\varepsilon C(t)}}
  \exp\left[-\frac{B(t)}{2 \varepsilon C(t)}(x-X_u(t))^2\right].
\end{gather*}
Conversely, the action of the left-inverse operator
\eqref{shapovalov:NL-EVOL-1} on the function
\eqref{shapovalov:GAUSS-1D-2} in the one-dimensional case gives
the function \eqref{shapovalov:GAUSS-1D-4}:
\begin{gather*}
\widehat U^{-1}(t,s,u)(\vec
x)=\int_{-\infty}^{+\infty}\sqrt{\frac{M_1(s,t)}{2\pi\varepsilon
  M_2(s,t)}}\\
  \phantom{\widehat U^{-1}(t,s,u)(\vec x)=}{}\times \exp\left[-\frac{1}{2\varepsilon}\frac{M_1(s,t)}{M_2(s,t)}
  \big(x-X_{\gamma}-M_3(s,t)(y-X_u(t))\big)^2\right]\\
\phantom{\widehat U^{-1}(t,s,u)(\vec
x)=}{}\times\sqrt{\frac{B(t)}{2\pi\varepsilon C(t)}}
  \exp\left[-\frac{B(t)}{2\varepsilon C(t)}(y-X_u(t))^2\right]dy\\
\phantom{\widehat U^{-1}(t,s,u)(\vec x)}{}
  =
  \sqrt{\frac{B_0}{2\pi\varepsilon C_0}}\exp\left[-
  \frac{B_0}{2\varepsilon C_0}(x-X_{\gamma})^2\right].
\end{gather*}
Because the solution of the nonlinear equation
\eqref{shapovalov:FPKE-1} is reduced to seeking the  solution of
linear equation \eqref{shapovalov:LIN-EQ-1} in terms of  the
moment $\vec X_u(t)$~\eqref{shapovalov:EE}, the symmetry operators
of these two equations are closely connected.

Equation \eqref{shapovalov:LIN-EQ-1} with the operator $\widehat
L$ having the form of \eqref{shapovalov:LIN-EQ-2} is a  special
case of the  linear evolution equation quadratic in derivatives
$\partial_{\vec x}$ and independent variables $\vec x$. This
equation is known to be integrated in  explicit form (see, e.g.,
\cite{shapovalov:MANKO}) which in turn leads to integrability of
the nonlinear FPKE \eqref{shapovalov:FPKE-1},
\eqref{shapovalov:FPKE-1a} according to \eqref{shapovalov:W} or
\eqref{shapovalov:GREEN-NLIN}. The basis of solutions and the
Green's function  for equation \eqref{shapovalov:LIN-EQ-1} can be
constructed with the help of symmetry operators of special form
following, for example, \cite{shapovalov:MANKO,
shapovalov:Popov,shapovalov:MANKO-Dodon75}.

Consider the symmetry operators for   \eqref{shapovalov:LIN-EQ-1}
and \eqref{shapovalov:FPKE-1}.

\section{The symmetry operators}
\label{shapovalov:SEC-SYMM}

The symmetry operators for equation \eqref{shapovalov:FPK-1} can
be found in various ways following the general ideas of symmetry
analysis \cite{shapovalov:MANKO,shapovalov:OVSIANNIKOV,
shapovalov:IBRAGIMOV, shapovalov:OLVER, shapovalov:FUSCH-SS}.

\subsection{The determining equation and intertwining}

Let us construct for a  function $\gamma(\vec x)$ of the space
$\mathcal{S}$ the function $u(\vec x,t)$ of \eqref{shapovalov:W}
using the solutions of the Cauchy problems \eqref{shapovalov:EE},
\eqref{shapovalov:CAUCHY-X} for the vector $\vec X_u(t)$ and
\eqref{shapovalov:LIN-EQ-1}, \eqref{shapovalov:LIN-EQ-3} for
$w(\vec x,t)$.

Let us take an operator $\widehat a(\vec x)$ acting in the space
$\mathcal{S}$ as the initial operator for a time depending
operator $\widehat A(\vec x,t)$:
\begin{equation*}
\widehat A(\vec x,0)=\widehat a(\vec x).
\end{equation*}
The function
\begin{equation}\label{shapovalov:SYMM-2}
\gamma_A(\vec x)=\frac{1}{\alpha_A} \widehat a(\vec x)\gamma(\vec
x),\qquad \alpha_A=\int_{\mathbb{R}^n} \widehat a(\vec
x)\gamma(\vec x)d\vec x.
\end{equation}
determines the vector $\vec X_{\gamma_A}$ by formula
\eqref{shapovalov:CAUCHY-X}, where $\gamma(\vec x)$ is replaced by
$\gamma_A(\vec x)$. Taking the vec\-tor~$\vec X_{\gamma_A}$ as the
initial condition for equation \eqref{shapovalov:EE}, we f\/ind
the vector  $\vec X_{u_A}(t)$.

Obviously, the operator  $\widehat A(\vec x, t)$  determined by
the conditions
\begin{gather}
\big( -\partial_t+\widehat H_{\rm nl}(\vec x,t,\vec X_{u_A}(t))
  \big)\widehat A(\vec x,t)=\widehat B(\vec x,t)\big(-\partial_t+
  \widehat H_{\rm nl}(\vec x,t, \vec X_{u}(t))\big),\label{shapovalov:SYMM-3} \\
\widehat A(\vec x,0)= \widehat a(\vec x)\label{shapovalov:SYMM-3a}
\end{gather}
is a symmetry operator for equation \eqref{shapovalov:FPKE-1}.
Here $\widehat B(\vec x,t)$ is  an  operator such that $\widehat
B(\vec x,t)(0)=0$. This operator plays the part of a Lagrangian
multiplier and it is determined together with~$\widehat A(\vec x,
t)$.

Equation \eqref{shapovalov:SYMM-3} is the determining equation for
the symmetry operators of equation \eqref{shapovalov:FPKE-1}. In
general, \eqref{shapovalov:SYMM-3} is a nonlinear operator
equation. But, given the initial function $\gamma(\vec x)$ and the
initial operator $\widehat a(\vec x)$, we can f\/ind the vectors
$\vec X_{u}(t)$ and $\vec X_{u_A}(t)$ without solving the equation
\eqref{shapovalov:FPKE-1}. On substitution of these vectors in
\eqref{shapovalov:SYMM-3} the operators $\widehat H_{\rm nl}(\vec
x,t, \vec X_{u}(t))$ and $\widehat H_{\rm nl}(\vec x,t, \vec
X_{u_A}(t))$ become linear. Then we can assume that $\widehat
a(\vec x)$,  $\widehat A(\vec x,t)$, and $\widehat B(\vec x,t)$
are linear operators in \eqref{shapovalov:SYMM-3}.

Notice that in the general case the operators  $\widehat A$ and
$\widehat B$ depend on $\vec X_{u}(t)$, $\vec X_{u_A}(t)$, i.e.
\begin{equation*}
\widehat A=\widehat A(\vec x,t;\vec X_{u}(t),\vec X_{u_A}(t)),
\qquad \widehat B=\widehat B(\vec x,t;\vec X_{u}(t), \vec
X_{u_A}(t)).
\end{equation*}
If  $\widehat B(\vec x,t)=\widehat A(\vec x,t)$ then $\widehat
A(\vec x,t)$ is called the {\it  intertwining} operator for the
linear operators, satisfying the condition
\begin{gather}\label{shapovalov:SYMM-3b}
\big( -\partial_t+\widehat H_{\rm nl}(\vec x,t, \vec
X_{u_A}(t))\big) \widehat A(\vec x,t)=\widehat A(\vec x,t)\big(
-\partial_t+\widehat H_{\rm nl}
(\vec x,t, \vec X_{u}(t))\big), \\
 \widehat A(\vec x,0)=\widehat a(\vec x).
\label{shapovalov:SYMM-3c}
\end{gather}

If  the operator $\widehat a(\vec x)$ is given in
\eqref{shapovalov:SYMM-3}, \eqref{shapovalov:SYMM-3a} (or  in
\eqref{shapovalov:SYMM-3b}, \eqref{shapovalov:SYMM-3c}),  these
conditions are the Cauchy problems determining the
operator~$\widehat A(\vec x,t)$.

Now, let $u(\vec x,t)$ be a solution of the Cauchy problem
\eqref{shapovalov:FPKE-1}, \eqref{shapovalov:FPKE-1a},
\eqref{shapovalov:CAUCHY}, and the operator~$\widehat A$ is
determined by the solution of the Cauchy problem
\eqref{shapovalov:SYMM-3}, \eqref{shapovalov:SYMM-3a} or
\eqref{shapovalov:SYMM-3b}, \eqref{shapovalov:SYMM-3c}. Then we
immediately obtain that the function
\begin{equation}\label{shapovalov:SYMM-OPERAT}
u_{A}(\vec x,t)=\widehat A(\vec x,t;\vec X_{u}(t),\vec X_{u_A}
(t))u(\vec x,t)
\end{equation}
is a solution of the Cauchy problem
\begin{gather*}
 \partial_t u_{A}(\vec x,t)=\widehat H_{\rm nl}(\vec x,t, \vec
X_{u_A}(t)) u_{A}(\vec x,t),\\
u_{A}(\vec x,0)=\gamma_{A} (\vec x)=\frac{1}{\alpha_A} \widehat
a(\vec x)\gamma(\vec x). 
\end{gather*}

Therefore, the operator $\widehat A(\vec x,t;\vec X_{u}(t), \vec
X_{u_A}(t))$ is  a    symmetry operator of the nonlinear FPKE
\eqref{shapovalov:FPKE-1}, \eqref{shapovalov:FPKE-1a}.

Notice that the operator $\widehat A(\vec x,t; \vec X_{u}(t), \vec
X_{u_A}(t))$ in \eqref{shapovalov:SYMM-OPERAT} is nonlinear due to
the presence of the vectors $\vec X_{u}(t)$, and $\vec
X_{u_A}(t)$.

\subsection{Symmetry operators of nonlinear and linear FPKE}

We now  deduce a relation connecting two solutions of the
nonlinear FPKE \eqref{shapovalov:FPKE-1},
\eqref{shapovalov:FPKE-1a} using a~symmetry operator of the linear
equation \eqref{shapovalov:LIN-EQ-1}, \eqref{shapovalov:LIN-EQ-2}.
This relation can be considered  a~symmetry operator for the
nonlinear FPKE.

To this end consider equation \eqref{shapovalov:W} which connects
the nonlinear Cauchy problem \eqref{shapovalov:CAUCHY} with the
linear Cauchy problem \eqref{shapovalov:LIN-EQ-3}.

Let $\widehat A(\vec x,t)$ be a symmetry operator of the  linear
equation \eqref{shapovalov:LIN-EQ-1}. Then the function
\begin{equation*}
w_A(\vec x,t)=\frac{1}{\widetilde{\alpha}_A}\widehat A(\vec
x,t)w(\vec x,t), \qquad \widetilde{\alpha}_A=\int_{{\mathbb
R}^n}\widehat A(\vec x,0)w(\vec x,0)d\vec x,
\end{equation*}
is another solution of  the linear equation, which is  determined
by \eqref{shapovalov:LIN-EQ-1}, \eqref{shapovalov:LIN-EQ-2}. For
$t=0$ we have
\begin{equation*}
w_A(\vec x,0)=\frac{1}{\widetilde{\alpha}_A}\widehat A(\vec x,0)
\tilde\gamma(\vec x)\equiv \gamma_A(\vec x)
\end{equation*}
and
\begin{equation*}
\int_{{\mathbb R}^n} \gamma_A(\vec x)d\vec x=1
\end{equation*}
which leads to normalization of the function $w_A(\vec x,t)$:
\begin{equation*}
\int_{{\mathbb R}^n} w_A(\vec x,t)d\vec x=1.
\end{equation*}
On the other hand, the function $w_A(\vec x,t)$ is not centered
for a symmetry operator  $\widehat A(\vec x,t)$ of general form.
In other words, for $t=0$ the vector
\begin{equation*}
\vec \lambda_A=\int_{{\mathbb R}^n}\vec x w_A(\vec x,0)d\vec x=
\int_{{\mathbb R}^n}\vec x\gamma_A(\vec x)d\vec x
\end{equation*}
is nonzero.

To construct a solution of the nonlinear equation
\eqref{shapovalov:FPKE-1}, which would  correspond to the
solu\-tion~$w_A(\vec x,t)$ of the linear equation
\eqref{shapovalov:LIN-EQ-1} with the use of relation
\eqref{shapovalov:W}, the function $w_A(\vec x,t)$, being a
solution of equation \eqref{shapovalov:LIN-EQ-1}, should  be
centered.

We can immediately check that  equation
\eqref{shapovalov:LIN-EQ-1} is invariant under  the change of
variables $t'=t$, $\vec x'=\vec x- \vec l(t) $, where $\vec l(t)$
satisf\/ies the condition $ \dot{\vec l}(t)=-\Lambda\vec l(t)$.

Taking into account this property, let us introduce a vector
\begin{equation*}
\vec l_A(t)=\int_{{\mathbb R}^n}\vec x w_A(\vec x,t)d\vec x
\end{equation*}
which satisf\/ies the Cauchy problem
\begin{gather*}
\dot{\vec l}_A(t)=-\Lambda \vec l_A(t),\qquad
 \vec l_A(0)= \vec \lambda_A.
\end{gather*}
Then the function
\begin{equation*}
\tilde w_A(\vec x,t)=w_A(\vec x+\vec l_A(t),t)
\end{equation*}
satisf\/ies equation \eqref{shapovalov:LIN-EQ-1} and the initial
condition{\samepage
\begin{equation*}
\tilde{ w}_A(\vec x,0)=w_A(\vec x+\vec \lambda_A,0)= \gamma_A(\vec
x+\vec \lambda_A).
\end{equation*}
The function $\gamma_A(\vec x+\vec\lambda_A)$ is normalized and
centered. The same is true for $\tilde{w}_A(\vec x,t)$.}

Following \eqref{shapovalov:W},  we now construct a solution
$v_A(\vec x,t)$ of the nonlinear FPKE \eqref{shapovalov:FPKE-1}
related to $\tilde{w}_A(\vec x,t)$. Consider a vector $\vec Y(t)$
such that
\begin{equation*}
\dot{\vec Y}(t)=-(\Lambda+\varkappa K_3)\vec Y(t), \qquad \vec
Y(0)=\vec\lambda_A .
\end{equation*}

Immediate check shows that the function
\begin{equation*}
v_A(\vec x,t)=\tilde w_A(\vec x-\vec Y(t),t)
\end{equation*}
satisf\/ies the  equation
\begin{gather*}
\big\{-\partial_t+\varepsilon\Delta+\partial_{\vec x}
\big(\Lambda\vec x+\varkappa K_3\vec Y(t)\big)\big\}v_A(\vec
x,t)=0,\\
 v_A(\vec x,0)=\widetilde{w}_A(\vec x-\vec{\lambda}_A,0)=
\gamma_A(\vec x).
\end{gather*}
Notice that
\begin{gather*} \vec X_{v_A}(t)= \int_{{\mathbb R}^n}v_A(\vec x,t)\vec x d\vec x,
\qquad \vec X_{v_A}(0)=\int_{{\mathbb R}^n}\vec
x\gamma_A(\vec x)d\vec x=\vec{\lambda}_A,\\
\dot{\vec X}_{v_A}(t)=   -(\Lambda+\varkappa K_3)\vec X_{v_A}(t),
  \end{gather*}
then
\[
\vec Y(t)=\vec X_{v_A}(t).
\]
Therefore, $v_A(\vec x,t)$   satisf\/ies the nonlinear FPKE
\eqref{shapovalov:FPKE-1}. The relation between the solutions
$u(\vec x,t)$ and $v_A(\vec x,t)$  reads
\begin{gather*}
v_A(\vec x,t)=\tilde{w}_A(\vec x-\vec Y(t),t)=w_A(\vec x-\vec
Y(t)+\vec l_A(t),t)\\
\phantom{v_A(\vec x,t)}{}=\frac{1}{\widetilde{\alpha}_A}\widehat
A(\vec x-\vec Y(t)+\vec l_A(t),t)
  w(\vec x-\vec Y(t)+\vec l_A(t),t)\\
\phantom{v_A(\vec x,t)}{}=\frac{1}{\widetilde{\alpha}_A}\widehat
A(\vec x-\vec Y(t)+\vec l_A(t),t)   u(\vec x-\vec Y(t)+\vec
l_A(t)+\vec X_u(t),t).
\end{gather*}
This equation determines a symmetry operator $\widehat A_{\rm nl}$
of the nonlinear FPKE \eqref{shapovalov:FPKE-1}:
\begin{gather}
u_A(\vec x,t)=v_A(\vec x,t)\equiv \widehat A_{\rm nl}(\vec x,t)u(\vec x, t)\nonumber\\
\phantom{u_A(\vec x,t)}{}=\frac{1}{\alpha_A}\widehat A(\vec x-\vec
Y(t)+\vec l_A(t),t)
  u(\vec x-\vec Y(t)+\vec l_A(t)+\vec X_u(t),t).
\label{shapovalov:NL-SYMM-OPER}
\end{gather}

\subsection{Symmetry operators in terms of an operator Cauchy
problem}

Let us reformulate the construction of symmetry operators for the
nonlinear FPKE \eqref{shapovalov:FPKE-1},
\eqref{shapovalov:FPKE-1a} in terms of an operator Cauchy problem.

Consider the nonlinear Cauchy problem \eqref{shapovalov:FPKE-1},
\eqref{shapovalov:CAUCHY} associated  with the Cauchy problem
\eqref{shapovalov:EE}, \eqref{shapovalov:CAUCHY-X} for the vector
$\vec X_u(t)$ having the form of \eqref{shapovalov:VECTOR-X}.

With an operator
\begin{equation}\label{shapovalov:NL-SYMM-INIT-OPER}
\widehat a(\vec x):{\mathcal S}\rightarrow {\mathcal S}
\end{equation}
acting on  the initial function $\gamma(\vec x)\in \mathcal S$ of
the Cauchy problem \eqref{shapovalov:CAUCHY}, we def\/ine  a
function~$\gamma_A(\vec x)$ of the form (\ref{shapovalov:SYMM-2}),
which is taken as an initial condition for the Cauchy problem for
a~function~$u_A(\vec x,t)$
\begin{gather*}
\big\{-\partial_t +\widehat H_{\rm nl}(\vec x,t, \vec
X_{u_A}(t)) \big\}u_A(\vec x,t)=0, \nonumber\\ 
u_A(\vec x,0)=\gamma_A(\vec x),
\end{gather*}
where the vector $\vec X_{u_A}(t)$ is determined by the conditions
\begin{gather*}
\dot {\vec X}_{u_A}(t)=-(\Lambda +\varkappa K_3)\vec X_{u_A}(t),  \\
\vec X_{u_A}(0)=\vec X_{\gamma_A},\qquad \vec
X_{\gamma_A}=\int_{{\mathbb R}^n}\vec x\gamma_A(\vec
x)d\vec x. 
\end{gather*}
Notice that given the function $\gamma(\vec x)$  and the operator
$\widehat a(\vec x)$, we can f\/ind $\vec X_{u_A}(t)$ not finding
a~solution of the  FPKE \eqref{shapovalov:FPKE-1}.

We can immediately verify that the two functions
\begin{gather}
w(\vec x,t)=u(\vec x+\vec X_u(t),t), \label{shapovalov:NL-SYMM-INIT-01}\\
w_A(\vec x,t)=u_A(\vec x+\vec
X_{u_A}(t),t)\label{shapovalov:NL-SYMM-INIT-02}
\end{gather}
are solutions of the  linear equation \eqref{shapovalov:LIN-EQ-1}
and the initial conditions are
\begin{gather*}
 w(\vec x,0)=\gamma(\vec x+\vec X_\gamma),\qquad
 w_A(\vec x,0)=\gamma_A(\vec x+X_{\gamma_A}).
\end{gather*}
Def\/ine a linear operator $\widehat{\overline{A}}(\vec x,t)$ by
an operator equation
\begin{equation*}
\big[-\partial_t+\widehat L(\vec x,t),\widehat{\overline{A}} (\vec
x,t) \big]=0 
\end{equation*}
 with
the initial condition
\begin{gather*}
 \widehat{\overline{A}}(\vec x,0)=\widehat a(\vec x),
\end{gather*}
where $L(\vec x,t)$ is  def\/ined in \eqref{shapovalov:LIN-EQ-2}.
It can be shown that
\begin{gather*}
 \widehat{\overline{A}}(\vec x,t) w(\vec x,t)=\widehat
A(\vec x+\vec l_A(t),t) w(\vec x+\vec l_A(t),t).
\end{gather*}
Due to the uniqueness of the Cauchy problem solution, we have
\begin{equation*}
w_A(\vec x,t)=\widehat{\overline{A}}(\vec x,t)w(\vec x,t).
\end{equation*}

In view of \eqref{shapovalov:NL-SYMM-INIT-01},
\eqref{shapovalov:NL-SYMM-INIT-02} we have
\begin{equation*}
u_A(\vec x +\vec X_{u_A}(t),t)=\widehat{\overline{A}}(\vec x,t)
  u(\vec x+\vec X_u(t),t)
\end{equation*}
or
\begin{equation}
u_A(\vec x,t) =\widehat{\overline{A}}(\vec x-\vec X_{u_A}(t),t)
  u(\vec x-\vec X_{u_A}(t)+\vec X_u(t),t).\label{shapovalov:NL-SYMM-OPER-CAUCHY-5}
\end{equation}

This relation def\/ines a symmetry operator $\widehat
{\overline{A}}_{\rm nl}(\vec x,t)$
 of the nonlinear FPKE
\eqref{shapovalov:FPKE-1}:
\begin{equation*}
u_A(\vec x,t) =\widehat {\overline{A}}_{\rm nl}(\vec x,t) u(\vec
x,t). 
\end{equation*}

\subsection{Symmetry operators in terms of an evolution operator}

Using the evolution operator \eqref{shapovalov:NL-EVOL} and
left-inverse operator \eqref{shapovalov:NL-EVOL-1},  we can obtain
symmetry operators for the nonlinear FPKE
\eqref{shapovalov:FPKE-1}.

Let $\widehat a(\vec x)$ be an operator
\eqref{shapovalov:NL-SYMM-INIT-OPER} acting on an initial function
$\gamma(\vec x)$, and $u(\vec x,t)$ is the solution of the Cauchy
problem \eqref{shapovalov:FPKE-1},  \eqref{shapovalov:CAUCHY}.
Then the  function
\begin{equation}
u_A(\vec x,t)=\widehat{U}(t,s,\widehat{a}\,\widehat{U}^{-1}(t,s,
u))(\vec x) \label{shapovalov:NL-SYMM-EVOL}
\end{equation}
is a solution of the nonlinear FPKE corresponding the initial
function $\gamma_A(\vec x)$ of the form (\ref{shapovalov:SYMM-2}).
Equation \eqref{shapovalov:NL-SYMM-EVOL} def\/ines a symmetry
operator $\widehat{\widetilde{A}}_{\rm nl}$ for the nonlinear FPKE
\eqref{shapovalov:FPKE-1}:
\begin{equation}
u_A(\vec x,t)=\widehat {\widetilde{A}}_{\rm nl}(\vec x,t)u(\vec
x,t).\label{shapovalov:NNL-SYM}
\end{equation}

The one-dimensional case of \eqref{shapovalov:NL-SYMM-EVOL} reads
\begin{equation}
u_A(x,t)=\widehat{\widetilde{A}}_{\rm
nl}(x,t)u(x,t)=\widehat{U}(t,s, \widehat{a}\,\widehat{U}^{-1}(t,s,
u))(x),\label{shapovalov:NL-SYMM-EVOL1}
\end{equation}
where $\widehat U(t,s,\cdot)$ and $\widehat U^{-1}(t,s,\cdot)$ are
determined by \eqref{shapovalov:NL-EVOL} and
\eqref{shapovalov:NL-EVOL-1} in the one-dimensional case.

Consider an operator $\widehat{a}(x,t)$ of the form
\begin{gather*}
\widehat a(x,t)=M_1(t,s)(x-X_u(t))+(\varepsilon
  M_2(t,s)+M_3(t,s))\partial_x,\\
\widehat a(x,s)=x-X_{\gamma}+\partial_x,\qquad
X_{\gamma}=X_u(s).\nonumber
\end{gather*}
Here $M_1(t,s)$, $M_2(t,s)$, and $M_3(t,s)$ are solutions of the
system in variations \eqref{shapovalov:MATRICIANT} in the
one-dimensional case. Then for \eqref{shapovalov:NL-SYMM-EVOL1} we
have
\begin{gather*}
u_A(x,t)=\lim_{\tau\to t}\widehat U(t,s,[\partial_z+z-X_{\gamma}]
  \widehat U^{-1}({\tau},s,u))(x)\\
\phantom{u_A(x,t)}{}=\lim_{\tau\to t} \int_{-\infty}^{+\infty}dy
  \int_{-\infty}^{+\infty}dz
G_{\rm nl}(t,s,x,z,\gamma_A)[\partial_z+z-X_{\gamma}] G_{\rm
nl}^{-1}(\tau,s,z,y,u)u(y,\tau)\\
\phantom{u_A(x,t)}{}=\int_{-\infty}^{+\infty}dy
  \int_{-\infty}^{+\infty}dz\lim_{\tau\to t}\frac{1}{2\pi\varepsilon}
  \sqrt{\frac{M_1(t,s)M_1(s,\tau)}{M_2(t,s)M_2(s,\tau)}}\\
\phantom{u_A(x,t)=}{}\times\exp\Big\{-\frac{1}{2\varepsilon}
  \Big(x-X_{u_A}(t)-M_3(t,s)(z-X_{\gamma_A})\Big)^2
  \frac{M_1(t,s)}{M_2(t,s)}\Big\}[\partial_z+z-X_{\gamma}]\\
\phantom{u_A(x,t)=}{}\times\exp\Big\{-\frac{1}{2\varepsilon}\Big(z-X_{\gamma}
 -M_3(s,\tau)(y-X_u(\tau))\Big)^2
\frac{M_1(s,\tau)}{M_2(s,\tau)}\Big\}u(y,\tau)\\
\phantom{u_A(x,t)}{}=\big[M_1(t,s)(x-X_{u_A}(t)+M_3(t,s)X_{\gamma_A})-X_{\gamma}+(\varepsilon
  M_2(t,s)+M_3(t,s))\partial_x\big]\\
\phantom{u_A(x,t)=}{}\times u(x+X_u(t)-X_{u_A}(t)+M_3(t,s)(X_{\gamma_A}-X_{\gamma}),t)\\
\phantom{u_A(x,t)}{}=\widehat
a(x+X_u(t)-X_{u_A}(t)+M_3(t,s)(X_{\gamma_A}-X_{\gamma}),t)u
(x+X_u(t)-X_{u_A}(t)\\
\phantom{u_A(x,t)=}{}+M_3(t,s)(X_{\gamma_A}-X_{\gamma}),t).
\end{gather*}
Therefore, we have
\begin{gather}
u_A(x,t)=\widehat {\widetilde{A}}_{\rm nl}(x,t)u(x,t)=\widehat
a(x+X_u(t)-
  X_{u_A}(t)+M_3(t,s)(X_{\gamma_A}-X_{\gamma}),t)\nonumber\\
\phantom{u_A(x,t)=}{}\times u(x+X_u(t)-X_{u_A}(t)+M_3(t,s)
  (X_{\gamma_A}-X_{\gamma}),t).\label{shapovalov:LINSYM1}
\end{gather}
In calculating the symmetry operators we have used  the following
relations:{\samepage
\begin{gather*}
M_1(t,s)M_1(s,\tau)=M_1(t,\tau),\qquad
M_2(t,\tau)=M_1(s,\tau)M_2(t,s)+M_2(s,\tau)M_3(t,s),
\end{gather*}
which follow from \eqref{shapovalov:MATRICIANT} in the
one-dimensional case.}

Equality  \eqref{shapovalov:LINSYM1} determines a symmetry
operator in explicit form for equation
\eqref{shapovalov:GAUSS-1D-3}. This symmetry operator generates
solutions of equation \eqref{shapovalov:GAUSS-1D-3}. Let us
illustrate this by an example.

By acting  with  the operator \eqref{shapovalov:LINSYM1} on  the
function \eqref{shapovalov:GAUSS-1D-2}, we obtain
\begin{gather}
u_A(x,t)=\widehat a(x+X_u(t)-X_{u_A}(t)+M_3(t,s)
  (X_{\gamma_A}-X_{\gamma}),t)\times\nonumber\\
\phantom{u_A(x,t)=}{}\times u(x+X_u(t)-X_{u_A}(t)+M_3(t,s)
  (X_{\gamma_A}-X_{\gamma}),t)\nonumber\\
\phantom{u_A(x,t)}{}=\frac{1}{C(t)}\Big(C_0-\displaystyle\frac{1}{\varepsilon}{B_0}\Big)(x-X_{u_A}(t)+
  M_3(t,s)(X_{\gamma_A}-X_{\gamma})) \nonumber\\
\phantom{u_A(x,t)=}{}\times
u(x+X_u(t)-X_{u_A}(t)+M_3(t,s)(X_{\gamma_A}-X_{\gamma}),t).
\label{shapovalov:GAUSS-1D-3-1-1}
\end{gather}
The function $u_A(x,t)$ \eqref{shapovalov:GAUSS-1D-3-1-1} is a
solution of the equation
\begin{equation}
\{-\partial_t+ \varepsilon\partial_x^2+\partial_x\Lambda
x+\varkappa
K_3 X_{u_A}(t)\partial_x\}u_A(x,t)=0.\nonumber
\end{equation}
Notice that the symmetry operators  determined by
\eqref{shapovalov:NNL-SYM} are consistent with the  operators
\eqref{shapovalov:NL-SYMM-OPER} and
\eqref{shapovalov:NL-SYMM-OPER-CAUCHY-5} in the one-dimensional
case. Moreover, the  operator determined by
\eqref{shapovalov:LINSYM1} corresponds the operator determined by
relation \eqref{shapovalov:NL-SYMM-OPER}
\[
\widehat{\widetilde{A}}_{\rm {nl}}(\vec x,t)=\widehat A_{\rm
{nl}}(\vec x+\vec X_u(t),t),
\]
which follows from $ \dot
M_3(t,s)(X_{\gamma_A}-X_{\gamma})=-\Lambda
M_3(t,s)(X_{\gamma_A}-X_{\gamma}). $

\section{Conclusion}

Symmetry analysis of an equation is usually performed when
solutions of the equation are not known, and the basic purpose
consists in f\/inding as wide as possible  classes of partial
solutions by using the symmetries of the equation.

It should be noted that a direct calculation of symmetry operators
for a nonlinear equation is, as a rule,  dif\/f\/icult because of
the complexity of the determining equations
\cite{shapovalov:Pukhnachev}. The basic subject for study is the
symmetries related to the generators of one-parameter subgroups of
a~Lie group of symmetry operators \cite{shapovalov:OVSIANNIKOV}.
The determining equations for the symmetries are linear, and to
solve them, it is necessary to set the structure of the
symmetries.
Finding of
nonlocal symmetries for dif\/ferential equations or dif\/ferential
symmetries for nonlocal equations from the determining equations,
faces  mathematical problems.

In this  context, the algorithm proposed in this work to calculate
the symmetry operators for the FPKE  in explicit form is of
interest as it provides a possibility to consider  the properties
of symmetry operators for a nontrivial nonlinear equation. The
algorithm is stated in terms of a~direct nonlinear evolution
operator and the corresponding left-inverse operator, which
enables one to vary the structure of the obtained symmetry
operators.

As for the considered equation the evolution operator is found,
the symmetry operators are not of interest for f\/inding of
partial solutions of the equation. However, the symmetry operators
can be used to study the properties of the solutions obtained.
Furthermore the algebra of symmetry operators of the linear FPKE
(\ref{shapovalov:LIN-EQ-1}) (see, e.g.~\cite{shapovalov:LAGNO})
can be used to study the symmetry operators of the nonlinear
equation (\ref{shapovalov:FPKE-1}).

The developed approach  permits a generalization for FPKE's with a
smooth operator symbol of arbitrary form. In this case, we deal
with a solution of the FPKE approximate in the small parameter
$\varepsilon$ \cite{shapovalov:SHVEDOV}. The FPKE considered in
this work arises in constructing  the leading term of
semiclassical asymptotics for a FPKE with arbitrary
coef\/f\/icients.

\subsection*{Acknowledgements}
The work was supported by  President of the Russian Federation
Grant  No SS-5103.2006.2,  Deutsche Forschungsgemeinschaft. ROR
was supported in part by the scholarship of the nonprof\/it
Dynasty Foundation.

\pdfbookmark[1]{References}{ref}
\LastPageEnding

\begin{thebibliography}{99} 

\footnotesize\itemsep=0pt

\bibitem{shapovalov:Miller} Miller~W.,~Jr., Symmetry and separation of variables, Addison-Wesley, London~--
Amsterdam~-- Ontario~-- Tokio~-- New York,  1977.

\bibitem{shapovalov:MANKO} Malkin M.A., Manko V.I., Dynamic symmetries and coherent
states of quantum systems,  Nauka, Moscow, 1979 (in Russian).

\bibitem{shapovalov:OVSIANNIKOV} Ovsjannikov~L.V., Group analysis of dif\/ferential
equations, Nauka, Moscow, 1978 (English transl.: Academic Press,
New York,
  1982).

\bibitem{shapovalov:IBRAGIMOV}  Anderson~R.L.,  Ibragimov~N.H.,
 Lie--B\"acklund transformations in applications,  Philadelphia, SIAM, 1979.

\bibitem{shapovalov:OLVER} Olver~P.J., Application of Lie groups to
dif\/ferential equations, Springer, New York,  1986.

\bibitem{shapovalov:FUSCH-SS}
Fushchych~W.I.,  Shtelen~W.M.,  Serov~N.I., Symmetry analysis and
exact solutions of equations of nonlinear mathematical physics,
Kluwer, Dordrecht, 1993.

\bibitem{shapovalov:FUSCH-N} Fushchych~W.I., Nikitin~A.G.,
Symmetries of equations of quantum mechanics, Allerton
Press Inc., New York, 
 1994.

\bibitem{shapovalov:KRASIL-LYCHAG-VINIGR} Krasilshchik~I.S.,
Lychagin~V.V., Vinogradov~A.M., Geometry of jet spaces and
nonlinear partial dif\/ferential equations, {\it Advanced Studies
in Contemporary Mathematics}, Vol.~1, Gordon and Breach
Science Publishers, New York,  1986.

\bibitem{shapovalov:GAETA} Gaeta~G., Nonlinear symmetries and
nonlinear equations, {\it Mathematics and its Applications},
Vol.~299,  Kluwer Academic Publishers Group, Dordrecht, 1994.

\bibitem{shapovalov:BELUCH-Frank2005} Frank T.D.,
Nonlinear Fokker--Planck equations, Springer, Berlin, 2004.

\bibitem{shapovalov:BELUCH-TRIFONOV}  Bellucci~S.,  Trifonov~A.Yu.,
Semiclassically-concentrated solutions for the one-dimensional
Fokker--Planck equation with a nonlocal nonlinearity, {\it
J.~Phys.~A: Math. Gen.} {\bf 38} (2005),  L103--L114.

\bibitem{shapovalov:MASLOV2}
Maslov~V.P., The complex WKB method for nonlinear equations. I.
Linear theory,  Birkhauser Verlag, Basel~-- Boston~-- Berlin, 1994.

\bibitem{shapovalov:BEL-DOB} Belov~V.V., Dobrokhotov~S.Yu., 
Semiclassical Maslov asymptotics with complex phases.~I. General approach, {\it Teor.
Mat. Fiz.} {\bf 92} (1992), 215--254 (English transl.: {\it Theoret.
and Math. Phys.} {\bf 92} (1992), 843--868).

\bibitem{shapovalov:Dobrokh-Martin}  Dobrokhotov S.Yu., Martinez Olive V.,
Localized asymptotic solutions of the magneto dynamo equation in
ABC-f\/ields, {\it Mat. Zametki} {\bf 54} (1993), no.~4, 45--68 (English
transl.: {\it Math. Notes} {\bf 54} (1993), 1010--1025).

\bibitem{shapovalov:Albeverio}  Albeverio S., Dobrokhotov S.Yu., Poteryakhin M.,
On quasimodes of small dif\/fusion operators corresponding to
stable invariant tori with nonregular neighborhoods, {\it
Asymptot. Anal.} {\bf 43} (2005), no.~3, 171--203.

\bibitem{shapovalov:maslovv}  Maslov V.P.,
Global exponential asymptotic behavior of the solutions of
tunnel-type equations and the problem of large deviations, {\it
Trudy Mat. Inst. Steklov.} {\bf 193} (1984), 150--180.



\bibitem{shapovalov:BEL-SHAP-TRIF1} Belov V.V., Shapovalov A.V., Trifonov A.Yu.,
The trajectory-coherent approximation and the system of moments
for the Hartree type equation, {\it Int. J. Math. Math. Sci.} {\bf 32} (2002), 325--370,
\href{http://arxiv.org/abs/math-ph/0012046}{math-ph/0012046}.

\bibitem{shapovalov:BEL-SHAP-TRIF2}
Belov~V.V., Trifonov~A.Yu., Shapovalov~A.V., Semiclassical
trajectory-coherent approximations to Hartree type equations, {\it
Teor. Mat. Fiz.} {\bf 130} (2002), 460--492 (English transl.: {\it
Theoret. and Math. Phys.} {\bf 130} (2002), 391--418).


\bibitem{shapovalov:LIS-TRIF-SHAP1} Lisok A.L., Trifonov A.Yu., Shapovalov A.V., The
evolution operator of the Hartree-type equation with a~quadratic
potential, {\it J. Phys. A: Math. Gen.} {\bf 37} (2004), 4535--4556,
\href{http://arxiv.org/abs/math-ph/0312004}{math-ph/0312004}.

\bibitem{shapovalov:LIS-TRIF-SHAP3} Shapovalov~A.V., Trifonov~A.Yu., Lisok~A.L., Exact
solutions and symmetry operators for the nonlocal
Gross--Pitaevskii equation with quadratic potential, {\it SIGMA} {\bf 1} (2005), 007, 14 pages,
\href{http://arxiv.org/abs/math-ph/0511010}{\mbox{math-ph/0511010}}.

\bibitem{shapovalov:SHVEDOV} Shvedov O.Yu., Semiclasical symmetries
{\it Ann. Phys.} {\bf 296} (2002), 51--89.

\bibitem{shapovalov:TRIF2}
Trifonov A.Yu., Trifonova L.B., Fokker--Planck--Kolmogorov
equation's with a nonlocal nonlinearity in a~semiclassical
approximation, {\it Izv. Vyssh. Uchebn. Zaved. Fiz.} {\bf 45} (2002), no.~2, 
121--129 (English transl.: {\it Russian Phys. J.} {\bf 45} (2002), 118--128).

\bibitem{shapovalov:TRIF-REZAEV_izv} Bezverbny A.V., Gogolev A.S.,
Trifonov A.Yu., Rezaev R.O., Nonlinear Fokker--Planck--Kolmogorov
equation in the semiclassical coherent trajectory approximation
{\it Izv. Vyssh. Uchebn. Zaved. Fiz.} {\bf 48} (2005), no.~6, 38--47
(English transl.: {\it Russian Phys. J.} {\bf 48} (2005), 592--604).

\bibitem{shapovalov:SHIINO-2001} Shiino M., Yoshida K.,
Chaos-nonchaos phase transitions induced by external noise in
ensembles of nonli\-near\-ly coupled oscillators, {\it Phys. Rev. E} {\bf 63} (2001), 026210, 6~pages.

\bibitem{shapovalov:SHIINO-2003} Shiino M.,
Stability analysis of mean-f\/ield-type nonlinear Fokker--Planck
equations associated with a gene\-ralized entropy and its
application to the self-gravitating system, {\it Phys. Rev.~E} {\bf 67} (2003), 056118, 5~pages.

\bibitem{shapovalov:DROZDOV-MORILLO-1}
Drozdov A.N., Morillo M., Validity of basic concepts in nonlinear
cooperative Fokker--Plank models, {\it Phys. Rev. E} {\bf 54} (1996),
3304--3313.

\bibitem{shapovalov:SCHULLER-VOGT}
Schuller F.P., Vogt P., Product structure of heat phase space and
branching Brownian motion, {\it Ann. Phys.} {\bf 308} (2003),
528--554,
\href{http://arxiv.org/abs/math-ph/0209016}{math-ph/0209016}.

\bibitem{shapovalov:FRANK-2004}
Frank T.D., Classical Langevin  equations for the free electron
gas and blackbody radiation, {\it J. Phys. A: Math. Gen.} {\bf 37} (2004), 3561--3567.

\bibitem {shapovalov:TATARSKY} Tatarskii V.I., The Wigner representation of quantum mechanics,
{\it Uspekhi Fiz. Nauk} {\bf 139} (1983), 587--619 (English transl.:
{\it Soviet Phys. Uspekhi} {\bf 139} (1983), 311--327).

\bibitem{shapovalov:Popov} Popov M.M., Green functions for
the Schr\"odinger equation with a quadric potential, in Problems
of Mathe\-matical Physics, Vol.~6,  Leningrad, 1973, 119--125 (in
Russian).

\bibitem{shapovalov:MANKO-Dodon75} Dodonov V.V., Malkin I.A., Man'ko V.I.,
Integrals of motion, Green functions and coherent states of
dynamic systems, {\it  Internat. J. Theoret. Phys.} {\bf 14} (1975),
37--54.

\bibitem{shapovalov:Pukhnachev} Pukhnachov V.V.,
Transformations of equivalence and the hidden symmetry of
evolution equations, {\it Dokl. Akad. Nauk SSSR} {\bf 294} (1987),
535--538 (English transl.: {\it Sov. Math. Dokl.} {\bf 34} (1987),
555--558).

\bibitem{shapovalov:LAGNO} Lahno V.I., Spichak S.V., Stogniy V.I., Symmetry analysis of evolution
type equations,  Computer Research Institute, Moscow~-- Igevsk, 
2004 (in Russian).

\end{thebibliography}
\end{document}